\documentstyle[11pt]{article}
\newdimen\paperwidth
\newdimen\paperlength \newdimen\margin \newdimen\vmargin
= .75 truein \vmargin = 1 truein
 \textwidth=\paperwidth
\advance \textwidth by -2\margin \advance\hoffset by
-1truein \advance\hoffset by \margin
\textheight=\paperlength \advance \textheight by -2\vmargin
\advance\voffset by -1truein \advance\voffset by \vmargin
\advance\textheight by -2\baselineskip
\begin{document}

\begin{titlepage} \title{ {\bf  The Role of Boundary
Conditions  }\\
 {\bf in the Real-Space Renormalization Group
}\thanks{Work partly supported by CICYT under contracts
AEN93-0776 (M.A.M.-D.) and PB92-109 ,
 European Community Grant ERBCHRXCT920069 (G.S.).} }

\vspace{2cm}    \author{ {\bf Miguel A.
Mart\'{\i}n-Delgado}\dag \mbox{$\:$} and {\bf Germ\'an
Sierra}\ddag \\ \mbox{}    \\ \dag{\em Departamento de
F\'{\i}sica Te\'orica I}\\ {\em Universidad Complutense.
28040-Madrid, Spain }\\ \ddag{\em Instituto de
Matem\'aticas y F\'{\i}sica Fundamental. C.S.I.C.}\\ {\em
Serrano 123, 28006-Madrid, Spain } } \vspace{5cm}
\maketitle \def\baselinestretch{1.3} \begin{abstract}

We show that the failure of the real-space RG method in
the 1D tight-binding model is not  intrinsic to the method
as considered so far but depends on the choice of
boundary  conditions. For fixed BC's the failure does
happen. For free BC's we present a new analytical block
RG-method which gives the exact ground state  of the model
and the correct $1/N^2$-law for the energy of the first
excited state in the  large $N$(size)-limit. We also give
a  reconstruction method for the wave-functions of the
excited states.

\ \

\ \

\ \

\end{abstract}

\vspace{2cm} PACS numbers:  05.50.+q, 71.10.+x, 64.60.Ak

\vskip-17.0cm \rightline{UCM/CSIC-95-09} \rightline{{\bf
September 1995}} \vskip3in \end{titlepage}

\newpage
 \def\baselinestretch{1.5} \noindent

\section{Introduction}

Real space Renormalization Group (RG) methods originated
from the study of the Kondo problem by Wilson
\cite{wilson}.
 It was clear from the beginning that one could not hope
to achieve the accuracy  Wilson obtained for the Kondo
problem when dealing with more complicated many-body
quantum Hamiltonians. The key difference is  that in the
Kondo model there exists  a {\em recursion relation} for
Hamiltonians at each step  of the RG-elimination of
degrees of freedom. Squematically,

\begin{equation} H_{N+1} = H_N + \mbox{hopping boundary
term}         \label{a} \end{equation}

\begin{equation} H_{N+1} = R(H_N)        \label{b}
\end{equation}

\noindent where $R$ is the RG transformation.

\noindent The existence of such recursion relation
facilitates enormously the work, but as it  happens it is
specific of {\em impurity problems}.

{}From the numerical point of view, the Block
Renormalization Group (BRG)
 procedure proved to be not  fully reliable in the past
particularly in comparison with other numerical
approaches, such as  the Quantum MonteCarlo method which
were being developed at the same time. This was one of
the reasons why the BRG methods remained undeveloped
during the '80's until the begining of  the '90's when
they are making a comeback as one of the most powerful
numerical tools  when dealing with zero temperature
properties of many-body systems, a situation where the
Quantum MonteCarlo methods happen to be particularly badly
behaved as far as fermionic  systems is concerned
\cite{hirsch}.

As it happens,  the BRG gives a good qualitative picture
of many properties  exhibited by quantum lattice
Hamiltonians: Fixed points, RG-flow, phases of the system
etc. as well as good quantitative results for some
properties such as ground state energy and others
\cite{drell}, \cite{jullien},  \cite{hirschII},
\cite{rabin}. For a review on the Block RG method see
\cite{jullienlibro} and  chapter 11 of reference
\cite{jaitisi}.
 However in some important instances the BRG method is off
the correct values of critical exponents by a  sensible
amount.

The first advance in trying to understand the sometimes
bad numerical performance of the BRG methods came in the
understanding of the effect of {\em boundary conditions}
(BC) on the  standard RG procedure \cite{white-noack}.

White and Noack \cite{white-noack} pointed out that the
standard BRG approach of neglecting all  connections to
the neighbouring blocks during the diagonalization of the
block Hamiltonian $H_B$  introduces large errors which
cannot be corrected by any reasonable increase in the
number of  states kept. Moreover, in order to isolate the
origin of this problem they study an extremely simple
model: a free particle in a 1D lattice. As a matter of
fact, it was Wilson \cite{wilson-unp}  who pointed out the
importance of understanding real-space RG in the context
of this simple  tight-binding model where the standard BRG
clearly fails as we are going to show.

\noindent The reason for this failure can be traced back
to the importance of the boundary  conditions in
diagonalizing the states of a given block Hamiltonian
$H_B$ in which the lattice is  decomposed into.  Notice
that in this fashion we are isolating  a given block from
the rest of the  lattice and this applies a {\em
particular BC}  to the block. However, the block is not
truly  isolated! A statement which is the more relevant
the more strongly correlated  is the system  under
consideration. Thus, if the rest of the lattice were there
it would apply different  BC's to the  boundaries of the
block. This in turn makes the standard
block-diagonalization conceptually not  faithfully suited
to account for the interaction with the rest of the
lattice.

Once the origin of the problem is brought about the
solution is also apparent: devise a method to  change the
boundary conditions in the block in order to mimick the
interaction with the rest of  the lattice. This is called
the Combination of Boundary Conditions (CBC) method which
yields  very good numerical results. This method has not
yet been generalized to interacting systems. However in
reference \cite{white} an alternative approach is proposed
under the name of  Density Matrix Renormalization Group
(DMRG) which applies to more general situations and also
produces quite accurate results. Despite of the impresive
numerical accuracy of these methods there is not a clear
understanding of why they perform so well or whether they
would be valid in dimensions higher than one.

\section{Boundary Conditions and Real-Space RG}

In this letter we want to reconsider again the role of
BC's in the real space RG method for the  case of a
single-particle problem in a box. The continuum version of
this Hamiltonian is  simply $H = - \frac{\partial
^2}{\partial x^2}$. We shall consider open chains with two
types of  BC's at the ends:

\begin{equation} \mbox{Fixed BC's:} \ \ \ \  \psi (0) =
\psi (L) = 0           \label{1a} \end{equation}

\begin{equation} \mbox{Free BC's:}  \ \ \ \
\frac{\partial \psi}{\partial x} (0) =
 \frac{\partial \psi}{\partial x} (L) = 0
\label{1b} \end{equation}

\noindent The lattice version of $H$ for each type of BC's
is given as follows:

\begin{equation}
 H_{Fixed} = \left( \begin{array}{cccccc} 2 & -1 & & & & \\
 -1 & 2 & -1& &  & \\
 & -1 &2 & & &  \\ &  &  &  \ddots &  &  \\
 & &  & & 2 & -1  \\
 & &  & & -1 & 2       \label{2} \end{array}
\right) \end{equation}

\begin{equation}
 H_{Free} = \left( \begin{array}{cccccc} 1 & -1 & & & & \\
 -1 & 2 & -1& &  & \\
 & -1 &2 & & &  \\ &  &  &  \ddots &  &  \\
 & &  & & 2 & -1  \\
 & &  & & -1 & 1       \label{3} \end{array}
\right) \end{equation}

\noindent The only difference between $H_{Fixed}$ and
$H_{Free}$ appear at the first and last  diagonal entry
($2 \leftrightarrow 1$). The exact solution of (\ref{2})
and (\ref{3}) is very well-known  and we give it for
completeness:

\begin{equation} \mbox{Fixed BC's:} \ \ \ \  \psi_n (j) =
N_n^{Fx} \sin \frac{\pi (n+1)}{N+1} j,\ \  E_n =  4 \sin
^2(\frac{\pi (n+1)}{2(N+1)})       \label{4a}
\end{equation}

\begin{equation} \mbox{Free BC's:}  \ \ \ \  \psi_n (j) =
N_n^{Fr} \cos \frac{\pi n}{N} (j - \frac{1}{2}), \ \  E_n
=    4 \sin ^2(\frac{\pi n}{2N})      \label{4b}
\end{equation} \[ j = 1,2,\ldots ,N; \ \ \ \ n =
0,1,\ldots ,N-1. \]

\noindent where the $N_n's$ are normalization constants
and $N$ is the number of sites of the chain.

Before getting into the problem of the renormalization of
these Hamiltonians, it is worth to pointing  out another
physical realization of $H_{Free}$. Consider the following
Hamiltonian for the anisotropic Heisenberg model in an
open chain of $N$ sites,

\begin{equation} H_{\Delta } = -\frac{1}{2} \sum
_{j=1}^{N-1} (\sigma_j^x \sigma_{j+1}^x +  \sigma_j^y
\sigma_{j+1}^y + \Delta (\sigma_j^z \sigma_{j+1}^z -
1))    \label{x1} \end{equation}

\noindent where $\Delta =+1$ ($\Delta =-1$) corresponds to
the Ferromagnetic case  (Antiferromagnetic case).

\noindent When $\Delta >1$, the ground state of $H_{\Delta
} $ is the Ferromagnetic state  $|F\rangle$ given by:

\begin{equation} |F\rangle = |\uparrow \uparrow \ldots
\uparrow \rangle  \label{x2} \end{equation}

\begin{equation} H_{\Delta } |F\rangle = 0  \label{x3}
\end{equation}

\noindent Now, let us denote by $|x\rangle $ the state
which is obtained from the Ferromagnetic  state
$|F\rangle$ when one spin is overturned at the position
$x$ of the chain. The subspace  spanned by the $|x\rangle
$ states is the 1 magnon subspace of the Ferromagnetic
model. It is readly checked that the restriction of the
$H_{\Delta } $ Ferromagnetic Heisenberg Hamiltonian takes
the following form:

\begin{equation}
 H_{Magnon}(\Delta ) = \left( \begin{array}{cccccc} 1 & -1
& & & & \\
 -1 & 2 \Delta & -1& &  & \\
 & -1 &2 \Delta  & & &  \\ &  &  &  \ddots &  &  \\
 & &  & & 2 \Delta & -1  \\
 & &  & & -1 & 1       \label{x4} \end{array}
\right) \end{equation}

\noindent If $\Delta =+1$ it precisely coincides with
$H_{Free}$:

\begin{equation}
 H_{Magnon}(\Delta =+1)  = H_{Free}  \label{x5}
\end{equation}

\noindent Thus we arrive at the following mapping: {\em A
simple magnon above a ferromagnetic background satisfies
Free BC's}. We shall come back to this connection at the
conclusions for  it opens the door to the introduction of
interactions in the model.

Now let us get to the problem of renormalizing the
tight-binding Hamiltonians (\ref{2})-(\ref{3}).

In figures 1 and 2 we show the ground state and first
excited states of the chain with fixed and  free BC's.

It is clear from Fig.1 that {\em a standard Block RG
method is not appropiate  to study the ground state  of
fixed BC's since this state is non-homogeneous while the
block truncation does not take into account this fact.}

\noindent Each piece of the ground state within each block
satisfies BC's which vary from block to  block. This is
the motivation of reference \cite{white-noack} to consider
different BC's  in the  block method, yielding quite
accurate results as can be seen from Table 1 for
comparison of the  CBC method and the standard method.

\noindent We observe that the standard RG method performs
rather poorly as compared to the  CBC method which yields
quite the exact results.

\noindent The other alternative to the CBC method is the
Density Matrix RG method which can  be phrased by saying
that the rest of the chain produces on every block the
appropiate BC's  to be applied to its ends, and it has the
virtue that can be generalized to other models,  something
which is not the case as for the CBC method.

\noindent On the other hand, the ground state of
$H_{Free}$ is an homogeneous state (see Fig.2)  which in
turn suggests that a standard RG analysis may work for
this type of BC's. We shall  show that this is indeed the
case if the RG procedure is properly defined. The key of
our  RG-prescription is to notice that $H_{Free}$ has a
geometrical meaning: {\em $H_{Free}$ is the incidence
matrix of the graph in Fig.3}, and it is called minus the
discrete laplacian $-\Delta$ of that graph. Notice that
$H_{Fixed}$ has not such geometrical interpretation, in
fact, it concides with  the Dynkin diagram of the algebra
$A_N$.

\noindent Based on this observation the Kadanoff blocking
is nothing but the breaking of the graph  into $N/n_s$
disconnected graphs of $n_s$ sites each. We shall choose
$n_s = 3$ in our later  computation as shown in Fig. 4.

The previous geometrical interpretation of $H_{Free}$
suggests that we choose the block  Hamiltonian $H_B$ to be
the incidence matrix  of the disconnected graph in Fig.4,
namely,

\begin{equation}
 H_{B} = \left( \begin{array}{ccccccc} 1 & -1 & & & & & \\
 -1 & 2 & -1& &  &  & \\
 & -1 &1 & & &  & \\ &  &  &  1 & -1 &  & \\
 & &  &-1 & 2 & -1 &  \\
 & &  & & -1 & 1     &  \\
 & &  & &  &    & \ddots
 \label{5} \end{array}              \right) \end{equation}

\noindent in which case the interblock Hamiltonian
$H_{BB}$ which describes the interaction  between blocks
becomes:

\begin{equation}
 H_{BB} = \left( \begin{array}{cccccccc} 0 &  & & & & & &\\
  & 0 & & &  &  & &\\
 &  &1 & -1& &  & &\\ &  &  -1&  1 &  &  & &\\
 & &  &  & 0 &  &  &\\
 & &  & &  & 1     &  -1 &\\
 & &  & &  & -1   & 1 & \\
 & &  & &  &    &  &\ddots
 \label{6} \end{array}              \right) \end{equation}

\noindent $H_{BB}$ in turn also coincides with the
incidence matrix  of a graph which contains  the links
missing in Fig.4 (RHS) which connects consecutive blocks.
In a few words: our RG-prescription introduces free BC's
at the ends of every block. This condition fixes uniquely
the breaking of $H_{Free}$ into the sum $H_B + H_{BB}$.
This is the choice we make. It should be emphasized that
the splitting of $H_{Free}$ into two parts $H_B + H_{BB}$
is by no means unique, so that different choices may lead
to very different results.

\noindent Prior to any computation we notice that the
previous RG-prescription should lead to an  exact value of
the ground state energy, for the ground state of each
block is again a constant  function. The question is
therefore to what extent our method is capable of
describing the  excited states. We shall concentrate
ourselves to the first excited state since computations
can be carried out analytically.

First of all we diagonalize $H_B$ within each block of 3
sites,  keeping only the ground state $\psi _0^{(0)}$ and
the first excited state $\psi _1^{(0)}$ ($3 \rightarrow 2$
truncation).  The superscript denotes the initial step in
the truncation method. In the standard RG method we would
choose $\psi _0^{(0)}$  and $\psi _1^{(0)}$ as the
orthonormal basis for the truncated Hilbert  space and
obtain the effective Hamiltonian $H'_{B}$ and $H'_{BB}$.
In our case it is convenient to  express these effective
Hamiltonians in a basis expanded by the following linear
combination:

\begin{equation} \psi _+^{(0)}  = \frac{1}{\sqrt{2}} (\psi
_0^{(0)}  +  \psi _1^{(0)})                  \label{7a}
\end{equation}

\begin{equation} \psi _-^{(0)}  = \frac{1}{\sqrt{2}} (\psi
_0^{(0)}  -  \psi _1^{(0)})                  \label{7b}
\end{equation}

\noindent which are also an orthonormal basis of the
truncated Hilbert space.

\noindent In this basis the truncation of $H_B$ reads as
follows,

\begin{equation}
 H_B \longrightarrow H'_{B} = \left( \begin{array}{ccccc}
A & 0 & 0 & & \\
  0 & A & 0 & &  \\
 0 &  0 & A & & \\ &  &  &  \ddots &  \\
 & &  &  & A\\
 \label{8} \end{array}              \right) \end{equation}

\[  A = \frac{\epsilon}{2}   \left( \begin{array}{cc} 1 &
-1  \\
  -1 & 1   \\ \end{array}              \right) \]

\noindent with $\epsilon$ taking on the value
$\epsilon^{(0)} = 1$ in the initial step of the RG-method,
which is the energy of the  state $\psi _1^{(0)}$.

The truncation of $H_{BB}$ is more complicated, the result
being:

\begin{equation}
 H_{BB} \longrightarrow H'_{BB} = \left(
\begin{array}{cccccc} B & C & 0 & & & \\
  C^t & D + B & C & &  & \\
 0 &  C^t & D + B & & &  \\ &  &  &  \ddots &   & \\ &  &
&  D + B & C  \\
 & &  &  C^t & D \\
 \label{9} \end{array}              \right) \end{equation}

\[  B =   \left( \begin{array}{cc} a^2 & a b  \\
  a b & b^2   \\ \end{array}              \right) \ \  C
=   \left( \begin{array}{cc} -a b & -a^2  \\
  -b^2 & -a b   \\ \end{array}              \right) \ \  D
=   \left( \begin{array}{cc} b^2 & a b  \\
  a b & a^2   \\ \end{array}              \right) \]

\noindent  with $a$ $b$ taking on the values $a^{(0)} =
\frac{1}{\sqrt{6}} - \frac{1}{2}$ and  $b^{(0)} =
\frac{1}{\sqrt{6}} + \frac{1}{2}$ in the initial step of
the RG-method.

The nice feature about the basis (\ref{7a})-(\ref{7b})
 is that all rows and columns of (\ref{8}) and (\ref{9})
add up to zero, just like the original Hamiltonians
(\ref{5}) and (\ref{6}), implying that the  constant
vector is an eigenvector with zero eigenvalue of the
renormalized Hamiltonian!

We shall call $H_{N/3}(\epsilon, a, b)$ the sum of the
Hamiltonians (\ref{8}) and (\ref{9}) for  generic values
of $\epsilon$, $a$ and $b$. Next step in our RG-procedure
is to form blocks  of 4 states of the new Hamiltonian
$H_{N/3}(\epsilon, a, b)$ and truncating to the two
lowest  $\psi _0^{(1)}$ and  $\psi _1^{(1)}$ energy states
within each 4-block (4 $\rightarrow$ 2 truncation).  The
reason for this change in the number of sites per block
(from 3 to 4)
 is motivated by the form of $H'_{BB}$ in (\ref{9}) and
the fact that if we try to make a second step in the
RG-method with 3-blocks the method is doomed to failure
because the constant  state of Fig.2 would no longer be
the ground state.

\noindent Fortunately enough, with 4-blocks if we define
new states  $\psi _+^{(1)}$ and  $\psi _-^{(1)}$ in the
same form as we did in Eq.(\ref{2}), we obtain that  the
new effective Hamiltonian is obtained by a redefinition of
the parameters, namely,

\begin{equation} H_{N/3}(\epsilon, a, b)  \longrightarrow
H_{N/6}(\epsilon', a', b') \label{10a} \end{equation}

\begin{equation} \epsilon' = \frac{\epsilon}{2} + a^2 +b^2
- \Delta  \label{10b} \end{equation}

\begin{equation} a' = \frac{1}{2 \sqrt{2}} \left[ a + b -
\frac{a (a^2 - 3 b^2 + \Delta) + \frac{b \epsilon}{2}}
{\sqrt{\Delta (\Delta + a^2 - b^2)}}   \right]  \label{10c}
\end{equation}

\begin{equation} b' = \frac{1}{2 \sqrt{2}} \left[ a + b +
\frac{a (a^2 - 3 b^2 + \Delta) + \frac{b \epsilon}{2}}
{\sqrt{\Delta (\Delta + a^2 - b^2)}}   \right]  \label{10d}
\end{equation}

\begin{equation} \Delta \equiv \sqrt{(a^2 - b^2)^2 +
(\frac{\epsilon}{2}  - 2 a b)^2} \label{10e} \end{equation}

\noindent In this fashion, the constant state of Fig.2 is
again the ground state of the model and  moreover, upon
iteration of Eqs.(\ref{10a})-(\ref{10e}) there are no
level crossing among the excited  states. Otherwise stated
this means that the level structure of the block
Hamiltonian $H_B$ is preserved  under the action of our
BRG-method based upon the reduction from 4 to 2 states.

The energy $E_1(N)$ of the first excited state of a chain
with $N = 3 \times 2^m$ sites can be  obtained iterating
$m$ times Eqs.(\ref{10a})-(\ref{10e}):

\begin{equation} E_1(N=3 \times 2^m) \equiv
\epsilon^{(m)}  \label{11} \end{equation}

\noindent The initial data are given by:

\begin{equation} \epsilon^{(0)} = 1, \ \ a^{(0)} =
\frac{1}{\sqrt{6}} - \frac{1}{2},\ \   b^{(0)} =
\frac{1}{\sqrt{6}} + \frac{1}{2}            \label{12}
\end{equation}

\noindent In Table 2 we give our results for small and
large number of iterations. For low values  of $N$ the
deviation of $\epsilon ^{(m)}$ with respect to the exact
result is small. Recall that we are only keeping two
states in our RG-procedure, and that the ground state
energy is exactly zero by construction!. But what is more
interesting about these results in  Table 2 is that we are
able to obtain the correct size dependence, i.e., $1/N^2$
of $\epsilon ^{(m)}$. As a matter of fact, the energy of
the first excited state behaves for large $N$ as
(\ref{4b}):

\begin{equation} E_1^{(exact)}(N) \sim c_{exact}/N^2, \ \
\ \ \mbox{with} \ \ c_{exact} = \pi^2          \label{13}
\end{equation}

\noindent while our BRG-method gives,

\begin{equation} E_1^{(BRG)}(N) \sim c_{BRG}/N^2, \ \ \ \
\mbox{with} \ \ c_{BRG} = 12.6751         \label{14}
\end{equation}

The achievement of the $1/N^2$-law is a remarkable result
which in turn allows us to match the  correct order of
magnitude of the energy. For instance, for 10 iterations
our RG-method with 2 states kept  gives  the energy of
the  order of $10^{-6}$, which is precisely the same order
of magnitude as for the CBC method  (see Table 1) but with
8 states kept in the case of Fixed BC's. Recall that the
standard BRG  performs as bad as a $10^{-2}$ order of
magnitude.

\section{Wave-Function Reconstruction}

We may also wonder whether we are able to make a
reasonable picture of the first excited state
wave-function based upon our BRG-procedure when compared
with the exact form  depicted in Fig.2. As we are working
with a real-space realization of the renormalization
group method, this is something we have at hand. To do
this we need to perform a ``reconstruction"  of the
wave-function. This reconstruction amounts to plot the
form of our aproximate  wave-function in each and every
of the 3-blocks out of the $2^{m+1}$ in which the
original chain is decomposed into under the BRG-procedure.
Recall that in the initial step we  started out with
blocks of 3 states keeping the two lowest states
$\psi_0^{(0)}$ and $\psi_1^{(0)}$  ($3 \rightarrow 2$
truncation).  In the next step we make blocks of 4 states
keeping the two lowest  states $\psi_0^{(1)}$ and
$\psi_1^{(1)}$  ($4 \rightarrow 2$ truncation) and then we
perfom the iteration procedure over and over. As a  result
of this procedure we may express the two lowest wave
functions of the $m+1$-th step  in terms of those of the
previous $m$-th step by means of the following matricial
form:

\begin{equation} \left( \begin{array}{c} \psi_0^{(m+1)}  \\
  \psi_1^{(m+1)} \\ \end{array}              \right)  =
\frac{1}{\sqrt{2}}  \left( \begin{array}{cc} 1 & 0  \\
  \alpha_m & \beta_m  \\ \end{array}              \right)
\left( \begin{array}{c} \psi_0^{(m)}  \\
  \psi_1^{(m)} \\ \end{array}              \right)_L  +
 \frac{1}{\sqrt{2}}  \left( \begin{array}{cc} 1 & 0  \\
  -\alpha_m & \beta_m  \\ \end{array}
\right)  \left( \begin{array}{c} \psi_0^{(m)}  \\
  \psi_1^{(m)} \\ \end{array}              \right)_R
\label{15} \end{equation}

\noindent where the LHS of Eq. (\ref{15}) represents the
wave function of $3\times 2^{m+1}$ sites  while in the RHS
we have a left-wave-function of $3\times 2^{m}$ sites and
another  right-wave-function of $3\times 2^{m}$ sites, so
that everything squares. The parameters  appearing in Eq.
(\ref{15}) turn out to be given by:

\begin{equation}
 \alpha_m = \frac{(\frac{\epsilon_m}{2} - 2 a_m b_m) +
(a_m^2 - b_m^2 + \Delta_m)} {2 \sqrt{\Delta_m (\Delta_m +
a_m^2 - b_m^2)}}     \label{16} \end{equation}

\begin{equation}
 \beta_m = \frac{(\frac{\epsilon_m}{2} - 2 a_m b_m) -
(a_m^2 - b_m^2 + \Delta_m)} {2 \sqrt{\Delta_m (\Delta_m +
a_m^2 - b_m^2)}}     \label{17} \end{equation}

\noindent with $\Delta_m$ as in Eq.(\ref{10e}). Their
initial values are $\alpha_0=1/\sqrt{10}$ and
$\beta_0=3/\sqrt{10}$.

\noindent We may recast Eq. (\ref{15})  in more compact
form by writing:

\begin{equation}
 \Psi^{(m+1)} = L_m  \Psi^{(m)}_L  +   R_m  \Psi^{(m)}_R
\label{18} \end{equation}

\noindent where

\begin{equation}
 L_m  =  \frac{1}{\sqrt{2}}  \left( \begin{array}{cc} 1 &
0  \\
  \alpha_m & \beta_m  \\ \end{array}
\right)    \label{19} \end{equation}

\begin{equation}
 R_m  =  \frac{1}{\sqrt{2}}  \left( \begin{array}{cc} 1 &
0  \\
  -\alpha_m & \beta_m  \\ \end{array}
\right)    \label{20} \end{equation}

\noindent We may call Eq.(\ref{18}) the {\em
reconstruction equation}. This is the master equation
that when iterated ``downwards" (reconstruction) allows us
to obtain the picture of our  approximate
BRG-wave-function corresponding to every and each block of
3 sites of the  $2^{m+1}$ blocks in which the chain is
decomposed into. At the end of the iteration procedure we
end up with expressions for the values of the 3-sites
wave-functions in terms of the  initial two lowest states
$\psi_0^{(0)}$ and $\psi_1^{(0)}$. The first one is a
constant function  while the second is a straight line of
negative slope. Thus, these two states turn out to be the
building blocks of our BRG-procedure. As a matter of
illustration, we present in Fig.5 the  plot for the
reconstruction of the first excited wave-function for a
chain of $N=48$ (m=3) sites long. We can observe that our
BRG-wave-function has the shape of a broken line around
the  exact form of the wave function. We have checked that
this broken shape remains when increasing  the length of
the chain so that its nature must be due to the fact that
we are only keeping  2 lowest states in our procedure. It
is expectable that upon increasing this number the  shape
of the aproximate wave function must be smoothed out.
Nevertheless, the important point is that our BRG-method
preserves the  number of nodes of the wave-function which
one of the defining features of the exact result.

When using the reconstruction equation to obtain the wave
function we may use a binary code  based upon the labels L
(left) and R (right) to keep track of the different
3-sites blocks which  make up the chain. Thus, in one
dimension the RG-blocks are in a one-to-one
correspondence   with a binary numerical system. In
general, for other dimensions we may state  squematically
the following  correspondence:

\[ \mbox{BRG-prescription} \longleftrightarrow
\mbox{``Number System"} \]

\section{Conclusions and Outlook}

In this letter we have shown that it is possible to
perform a successful real-space block RG  treatment of the
1D tight-binding model. In doing so we have elucidated the
role played by the  boundary conditions in the real-space
RG method for we have shown that the failure of the
BRG-method pointed out in \cite{white-noack} can be traced
back to a particular choice  of boundary conditions: if we
choose free BC's instead of fixed BC's the ground state
turns out  to be homogeneous, a property which is
necessary in order
 to have a suitable implementation of a blocking
procedure. Otherwise, for fixed BC's the blocks of the
RG-method cannot have fixed  BC's as well for in that case
one introduces anomalous behaviour in the ground state
wave-funcion which ruins the method. Thus, the failure of
the BRG-method accepted so far is not intrinsic to  the
method but depends on the choice of the boundary
conditions.

\noindent We have devised a BRG-method which always yields
the correct ground state of the  Hamiltonian with free
BC's for open chains of any length. As for the excited
states, our method is able to reproduce the correct
$1/N^2$-law exhibited by the exact first excited  state
(\ref{13})-(\ref{14}) in the large $N$-limit with the
correct order of magnitud. Since we  only keep 2 states -
to make computations analytical - at every step of our
truncation procedure, this means that we have been able to
capture the essential physics of the model in the most
economic way. For this reason we believe that our method
should also be able to account for the  correct properties
of the rest of excited states of the model.

\noindent We have given another physical interpretation
for the tight-binding Hamiltonian with  free BC's as the
restriction of the Ferromagnetic Heisenberg model to the
subspace of  one-magnon solutions (\ref{x5}). In the light
of this mapping we may interpret the $1/N^2$-law of our
BRG-method as in good agreement with the correct {\em
quadratic} dispersion relation  exhibited by the
Ferromagnetic magnons $E(k) \sim k^2 \sim 1/N^2$ for low
wave number $k$. Thus, from this point of view, our
RG-procedure is aplicable to the Ferromagnetic regime of
the Heisenberg model and we may wonder whether we may
extend it to the Antiferromagnetic  regime and obtain the
{\em linear} dispersion relation $E(k) \sim k \sim 1/N$.

\noindent Our BRG-method also gives a good qualitative
picture of the ground state and  first excited states
wave-functions in coordinate space as shown in Fig.5. In
fact, the number of  nodes is preserved by the
BRG-procedure. This good performance is due to a
reasonable choice  of the 2 lowest energy states kept in
the initial stage of the method. The reconstruction of
the  wave-function is based upon the master equation
(\ref{18}) which can be readly generalized to  account for
a bigger number of states kept.

\noindent Finally, we may address yet another
generalization of the present work by considering  a
Hamiltonian $H_{Free}(q)$ as in Eq.(\ref{3})
 but depending upon a complex parameter $q$. This
parameter enters in the matrix expression of the
Hamiltonian by replacing the diagonal  $(1,2,2, \ldots
,2,1)$ of Eq.(\ref{3})
 by $(q ,q + q^{-1},q + q^{-1}, \ldots  ,q +
q^{-1},q^{-1})$ while the off-diagonal terms remain as -1.
This corresponds to  a one-parameter deformation of the
boundary conditions in which the free case is recovered as
the  $q \rightarrow 1$ limit. Furthermore, it is more
interesting to consider this extension  from the point of
view of the Heisenberg-model connection for in this case
the Hamiltonian  comes with a boundary term derpending on
the $q$ parameter:

\begin{equation} H_{\lambda } = -\frac{1}{2} \sum
_{j=1}^{N-1} (\sigma_j^x \sigma_{j+1}^x +  \sigma_j^y
\sigma_{j+1}^y +  \frac{q + q^{-1}}{2} (\sigma_j^z
\sigma_{j+1}^z - 1)) -
  \frac{q - q^{-1}}{4} (\sigma_1^z  - \sigma_{N}^z)
\label{21} \end{equation}

\noindent The important point about this Hamiltonian is
that it is invariant under what is called the  quantum
group $SU_q(2)$.
 We have shown in \cite{q-germanyo} that this symmetry
plays an  important role in constructing an improved
version of the standard BRG-procedure. We call this  new
method a $q$-BRG-method and for instance it is able to
obtain correctly the line of  critical XXZ models, unlike
the standard BRG-method. This quantum symmetry must show
up in the free case when performing the previous
$q$-deformation.


\vspace{2cm}

{\em e-mail addresses}:  mardel@fis.ucm.es (M.A. M.-D.);
sierra@cc.csic.es (G.S.)

\newpage \section*{Table captions}

 {\bf Table 1 :} Exact, Standard RG and CBC Values of  Low
Lying  States for the 1D Tight-Binding  Model for a chain
of $N=2048$ sites.

{\bf Table 2 :} Exact and new standard RG values of the
first excited state for the 1D Tight-Binding Model.

\newpage \section*{Figure captions} \noindent

 {\bf Figure 1:} Ground state $\psi _0$ and first excited
state $\psi _1$ for the Hamiltonian  $H_{Fixed}$ with
fixed BC's.

 {\bf Figure 2 :}  Ground state $\psi _0$ and first
excited state $\psi _1$ for the Hamiltonian  $H_{Free}$
with fixed BC's.

 {\bf Figure 3 :} The one-dimensional open chain as a
graph.

 {\bf Figures 4 :}  Lattice decomposition into blocks of
$n_s$ sites ($n_s=3$).

 {\bf Figure 5 :} The reconstruction of the first excited
wave-function for a chain of $N=48$ (m=3) sites long.

\newpage

\begin{table}[p] \centering \begin{tabular}{|c|c|c|c|}
\hline \hline
  \mbox{Energies} & \mbox{ Exact}    & \mbox{ Standard RG}
&  \mbox{CBC} \\  \hline \hline
 $E_0$& $2.351\times 10^{-6}$ & $1.9207\times 10^{-2}$  &
$2.351\times 10^{-6}$  \\ \hline $E_1$ & $9.403\times
10^{-6}$ & $1.9209\times 10^{-2} $ & $9.403\times
10^{-6}$  \\  \hline
 $E_2$ & $2.116\times 10^{-5}$ & $1.9214\times 10^{-2}$ &
$2.116\times 10^{-5}$ \\ \hline \hline \end{tabular}
\caption{Exact, Standard RG and CBC Values of  Low Lying
States for the 1D Tight-Binding  Model for a chain of
$N=2048$ sites.}  \end{table}

\begin{table}[p] \centering \begin{tabular}{|c|c|c|c|}
\hline \hline
  \mbox{m} & \mbox{N}    & $E_1^{(exact)}(N)$ &
$\epsilon^{(m)}$ \\  \hline \hline
 $0$& $3$ & $1$  & $1$  \\ \hline $1$ & $6$ & 0.2679
&0.2792   \\  \hline $2$ & $12$& 0.0681 &  0.0765  \\
\hline $10$ &$3072$ & $1.0458 \times 10^{-6} $ &
$1.3421\times 10^{-6}$  \\  \hline $20$ & $\sim 3 \times
10^{6}$ & $ 9.9737 \times 10^{-13}$ & $1.2809 \times
10^{-12}$  \\  \hline $32$ & $\sim  10^{10}$ & $5.9448
\times 10^{-20}$ & $7.6347 \times 10^{-20}$    \\  \hline
  & $\gg 1$ & $\pi^2/N^2$ & $12.6751/N^2$ \\ \hline \hline
\end{tabular} \caption{Exact  and new standard RG values
of the first excited state
 for the 1D Tight-Binding  Model.}  \end{table}

\end{document}